\documentclass[prb,twocolumn,superscriptaddress,showpacs,preprintnumbers,amsmath,unsortedaddress,amssymb]{revtex4}

\usepackage[dvips]{graphicx}
\usepackage{bm}
\DeclareGraphicsExtensions{.eps}
\begin{document}
\newcommand{\dyfeal}{DyFe$_4$Al$_8$}
\newcommand{\minustau}{$(1-\tau,1-\tau,0)$}
\newcommand{\minustwotau}{$(1-2\tau,1-2\tau,0)$}
\newcommand{\fivetau}{$(5\tau,5\tau,0)$}
\newcommand{\minusthreetau}{$(1-3\tau,1-3\tau,0)$}
\newcommand{\threetau}{$(3\tau,3\tau,0)$}

\title{Resonant soft x-ray magnetic scattering from the $4f$ and $3d$ electrons in DyFe$_4$Al$_8$}

\author{T. A. W. Beale} 
\affiliation{Department of Physics, University of Durham, Rochester
Building, South Road, Durham, DH1 3LE, UK}

\author{S. B. Wilkins} \affiliation{Brookhaven National Laboratory, Condensed Matter Physics
  \& Materials Science Department, Upton, NY, 11973-5000, USA}
\author{P. D. Hatton}\email{p.d.hatton@dur.ac.uk}
\affiliation{Department of Physics, University of Durham, Rochester
Building, South Road, Durham, DH1 3LE, UK}

\author{P. Abbamonte}
\affiliation{Frederick Seitz Materials Research Laboratory, Department of Physics, University of Illinois, 1110 West Green St., Urbana, IL 61801-3080, USA }

\author{S. Stanescu}
\affiliation{European Synchrotron Radiation
  Facility, Bo\^\i te Postal 220, F-38043 Grenoble Cedex, France}
\altaffiliation[Present Address: ]{Synchrotron SOLEIL, l'Orme des Merisiers, Saint-Aubin BP 48, 91192 Gif-sur-Yvette, France}
\author{J. A. Paix\~{a}o}
\affiliation{Departamento de F\'{i}sica, University of Coimbra, P-3000 Coimbra, 
Portugal}

\date{\today}
\begin{abstract}
Soft x-ray resonant scattering has been used to examine the charge and magnetic interactions in the cycloidal antiferromagnetic compound \dyfeal. By tuning to the Dy $M_4$ and $M_5$ absorption edges and the Fe $L_2$ and $L_3$ absorption edges we can directly observe the behavior of the Dy $4f$ and Fe $3d$ electron shells.  Magnetic satellites surrounding the (110) Bragg peak were observed below 60 K.   The diffraction peaks display a complex spectra at the Dy $M_5$ edge, indicative of a split $4f$ electron band.  This is in contrast to a simple resonance observed at the Fe $L_3$ absorption edge, which probes the Fe $3d$ electron shell. Temperature dependant measurements detail the ordering of the magnetic moments on both the iron and the dysprosium antiferromagnetic cycloids.   The ratio between the intensities of the Dy $M_4$ and $M_5$ remained constant throughout the temperature range, in contrast to a previous study conducted at the Dy $L_{2,3}$ edges.  Our results demonstrate the ability of soft x-ray diffraction to separate the individual magnetic components in complicated multi-element magnetic structures.

\end{abstract}
\maketitle

\section{Introduction}

The combination of both a rare earth and a transition metal within a single compound raises interesting possibilities for studying their magnetic interactions\cite{brookes:5861,plugaru:134419,miguel-soriano:5884,rueff:12271,kou:8254,cooper:1095}.   Deciphering the individual magnetic contributions within these compounds is essential to obtain an overall understanding of the bulk properties of the sample.  The series of compounds $R\mathrm{Fe}_4\mathrm{Al}_8$ where $R$ is a rare earth or actinide element show unusual magnetic properties\cite{felner:951,gubbens:61,gal:745}.  The highly symmetric structure and the relatively weak interaction between the two magnetic lattices in the compound provides an ideal platform for studying the interplay between the magnetic ions.   The combined contribution from both the iron and the rare earth atoms to the overall magnetic structure of \dyfeal\  has stimulated considerable interest\cite{angst:174407,duong:020408,schafer:205,hagmusa:298}.   In addition,  compounds with a higher iron content, such as SmFe$_{10}$Si$_2$\cite{suski:43}, are ferromagnets with a high T$_C$ and have a relatively large magnetic anisotropy, factors that are technologically important for high field magnets.

\begin{figure}
\includegraphics[width=0.9\columnwidth]{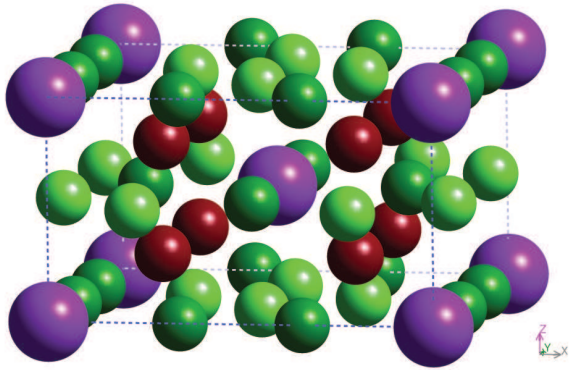}
\caption{\label{crystal}(color online) The crystal structure of \dyfeal. The dysprosium (purple) atoms are at the origin and the center of the unit cell in a cage formed by the 8 iron (brown) atoms. The remaining atomic positions are those of aluminium (green). }
\end{figure}

The compound forms a body centred unit cell $I4/mmm$~\cite{buschow:921}, with the dysprosium atoms occupying the eight corners and the central position (Fig.~\ref{crystal}).  Iron atoms occupy a further eight sites forming a cuboids around each dysprosium atom, with the aluminium atoms occupying the remaining positions.  Detailed studies using neutron diffraction\cite{paixao:6176} and x-ray magnetic scattering at the Dy $L_{2,3}$ edges\cite{langridge:2187} proposed a complicated magnetic structure, where at low temperatures both the dysprosium and iron magnetic moments are ordered into a cycloidal structure.   Below T$_N$ the iron orders antiferromagnetically in the [110] direction, with a small, long wavelength, modulation superimposed resulting in a cyclic rotation of the moments in the $ab$ plane with a period of approximately 15 unit cells.   At a lower temperature the dysprosium orders into the same form, and the same wavevector, only the cyclic rotation is counter-rotational to that of the iron\cite{paixao:6176}.

Paix\~{a}o~\emph{et al.}\cite{paixao:6176} using single crystal neutron diffraction and Langridge~\emph{et al.}\cite{langridge:2187} using resonant x-ray diffraction found weak superlattice peaks in the [110] directions around Bragg peaks, with a modulation of $n\tau$; where $\tau \approx 0.133$ and $n$ is odd.  These superlattice peaks arise from the long-range cycloidal antiferromagnetic ordering of the iron and the dysprosium atoms.  Paix\~{a}o \emph{et al.}\cite{paixao:6176} found the value of $\tau$ to vary with temperature.  Initially on cooling $\tau=0.138$ at T$_{N(\mathrm{Fe})}$, and then reducing to  $\tau=0.127$ at $\sim80$~K, before increasing again to $\tau=0.132$ at 2 K.  

Neutron studies are extremely sensitive to the magnetic moments on atoms, however they probe the total magnetic moment, and so it is difficult to determine the individual elemental contribution of the moments.   Although in the case of \dyfeal\ it was possible using neutron diffraction to separate the dysprosium moment from the overall magnetic moment through symmetry considerations, a direct probe to separate the elemental contribution to the total magnetic moment has, until recently not been available.  Resonant x-ray studies, such as those performed by Langridge \emph{et al.}, were restricted to looking at the $5d$ band in dysprosium, by tuning to the Dy $L_{2,3}$ edges.  These edges lie at approximately 8~keV and are easily accessible at resonant scattering beamlines.  Unfortunately the magnetic moments on the atoms in the title compound occur in the Fe $3d$ electron band, and the Dy $4f$ band, where the corresponding absorption edges lie in the region of 1~keV, energies which until very recently, have not been exploited using diffraction.

New advances in soft x-ray diffraction have allowed us to reach the $L$ edge of the transition metals~\cite{wilkins:187201} and the $M$ edges of the lanthanides~\cite{spencer:1725}.  We have discovered that the resonant enhancement of the magnetic x-ray scattering is considerably larger at these edges, and in addition these virtual transitions directly probe the electron bands of interest.  By tuning to different atomic absorption edges resonant x-ray scattering is thus an atomic selective, band sensitive, spectroscopic probe of the electronic and magnetic structure. For the first time we demonstrate that we can directly use resonant soft x-ray scattering at the individual atomic absorption edges to separate the individual magnetic ordering of the Dy $4f$ and Fe $3d$ electron bands.

Dramatic changes were found in the integrated intensity of the resonant (dipole) features at the Dy $L_3$ and $L_2$ edges by Langridge \emph{et al.}\cite{langridge:2187}.   This meant that the branching ratio (here defined as the ratio of the integrated intensities of the $L_3$ to $L_2$ edges) displayed a highly unusual temperature variation.   As we will demonstrate, our experiments which directly probe the Dy $4f$ electron band, show no such temperature variation in the $4f$ electron band, supporting the theory that the effect observed by Langridge \emph{et al.} occurs only in the $5d$ electron band, and appears entirely unconnected to the magnetic state of the system.

\section{Experimental Details and Results}

Experiments were performed on a single crystal of \dyfeal, grown by the Czochralski method from a levitated melt using high purity starting materials.  The sample was aligned with the [110] direction surface normal, presenting a highly polished 1~mm~$\times$~1~mm sample face to the beam. Previous neutron\cite{paixao:6176} and x-ray\cite{langridge:2187} scattering experiments on this sample have shown the existence of superlattice peaks, originating through the ordering of the iron and dysprosium sublattices.  The sample was examined at the X1B beamline, NSLS, and the ID08 beamline, ESRF.   Both beamlines provide incident x-rays with energies in the region of the Dy $M_{4,5}$ and Fe $L_{2,3}$ edges.  At each beamline the sample was mounted in the centre of an in-vacuum diffractometer, and cooled using a liquid helium cold finger allowing a base temperature of 18~K at X1B and 20~K at ID08. 
The APPLE II undulator x-ray source on the ID08 beamline supplied 100\% polarized light, which for this experiment was configured to give a horizontal linear polarized beam, which was coupled with a vertical scattering geometry.   By contrast, the X1B beamline has a 28\% vertically polarized beam at 1300 eV, while the scattering geometry was in the horizontal plane.

Extremely strong x-ray scattering was obtained from the (110) Bragg peak, which is observable at all energies.   The peak intensity was dramatically increased at x-ray energies close to the Dy $M_5$ edge.    The width of the peak corresponds to an x-ray penetration depth of the 5000$\mathrm{\AA}$.  This is approximately what would be expected from previous results\cite{wilkins:187201}, confirming that the Bragg peak resolution is limited through the penetration length rather than by sample quality.

\begin{figure}
\includegraphics[width=\columnwidth]{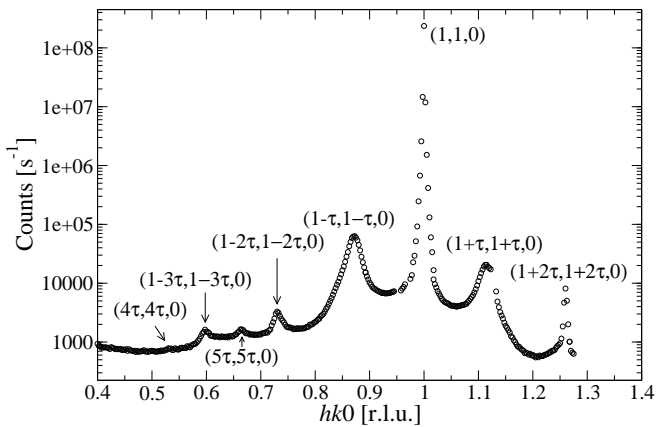}
\caption{\label{satellites} Superlattice peaks measured at the Dy $M_5$ edge in the [110] direction from the origin at 20~K.  Data measured at X1B. }
\end{figure}

Multiple satellites were found in the [110] direction around the (110) Bragg peak (Fig.~\ref{satellites}) with an incident x-ray energy corresponding to the the $M_5$ edge.  The scans in reciprocal space in the [110] direction correspond to the high resolution $\theta - 2\theta$ direction.   By comparison to the Bragg peak the correlation length of the magnetic superlattice peaks was $\sim250\mathrm{\AA}$, describing a significantly less correlated sublattice.

\begin{figure}
\includegraphics[width=0.8\columnwidth]{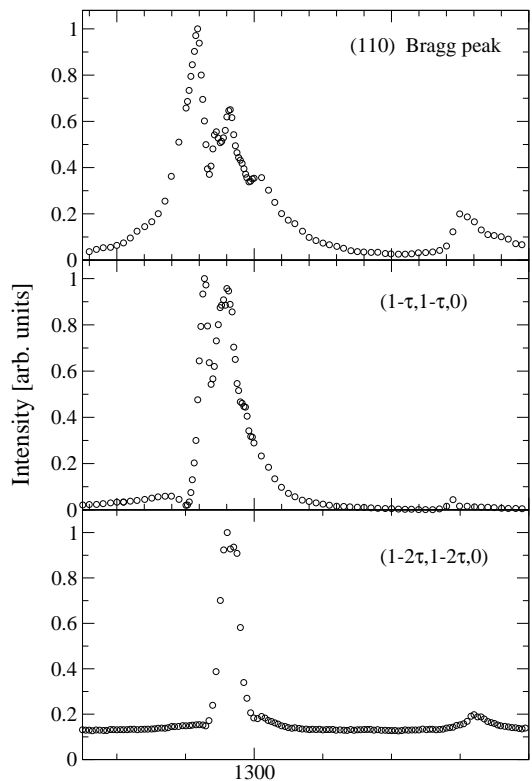}
\caption{\label{escans} Scattered x-ray intensity as a function of incident photon energy at the constant wavevector of (a) $\vec{Q}=(110)$ (b) $\vec{Q}=(1-\tau,1-\tau,0)$ and (c) $\vec{Q}=(1-2\tau,1-2\tau,0)$through the Dy $M_4$ and $M_5$ edges.  Data measured at 20~K ID08.}
\end{figure}

In addition to peaks appearing at $(1\pm n\tau, 1\pm n\tau,0)$, where $n$ is odd, corresponding to those observed by neutron scattering, we observed peaks where $n$ was even.  A high resolution energy scan at fixed $Q$ of the (1,1,0) Bragg peak was measured (Fig.~\ref{escans}a), and an energy scan was also collected at the \minustau\ satellite peak (Fig.~\ref{escans}b).  These energy scans covered the vicinity of the Dy $M_{4,5}$ edges, and were collected on the ID08 beamline at 20~K, with $\Delta E\approx50$ meV at 1300~eV.  The energy scans show a strong resonance at the Dy $M_5$ edge and a much weaker resonance at the $M_4$ edge.   The resonance of the satellite peak at the Dy $M_5$ edge was split into two strong peaks at 1293~eV and 1296~eV, with a shoulder on the high energy side of the latter.  The Bragg peak showed a similar resonance, but with a weaker resonant enhancement, and a much broader resonance.   Proportionally however, the $M_4$ edge is stronger, when compared to the $M_5$ edge, on the Bragg peak than the magnetic satellite peak.

A high resolution energy scan at fixed $Q$ was taken on the \minustwotau\ peak (Fig.~\ref{escans}(c)).   This spectrum is significantly different to spectra of either the Bragg peak of the magetic \minustau\ peak.  

The resonant x-ray scattering crossection  can be written in terms of geometrical factors depending on the incident and scattered photons and a tensor characteristic of the scattering ion. For dipole-dipole (E1) transitions there are three terms and the scattering amplitude is given as:

\begin{equation}
f^{E1} = (\vec{\epsilon^\prime}\cdot\vec{\epsilon})F^0 + 
[(\vec{\epsilon^\prime}\times\vec{\epsilon})\cdot\bm{\hat{z}}]F^1 + 
[\vec{\epsilon^\prime} \cdot \bm{\tilde{T}} \cdot \vec{\epsilon}]F^2,
\end{equation}
where $\vec{\epsilon}$ ($\vec{\epsilon^\prime}$) are the incident (scattered) x-ray's polarization vector and the terms $F^0$, $F^1$ and $F^2$  govern the strength of the resonance and are given by the atomic properties\cite{hannon:1245}.

The first term is a scalar and represents the resonant enhancement of the Thompson charge scattering and is responsible for the enhancement seen on the (110) Bragg reflection. This term does not result in any rotation of the photon's polarization.

The second term probes a tensor of rank 1 (vector) of odd time reversal symmetry, and results from a net spin polarization of the $4f$ states, a difference between overlap integrals or lifetime for the two channels\cite{hill:236,blume:3615,blume:1779}. This term results in a rotation of the photon's polarization. It is this term which is responsible for the magnetic satellite reflections corresponding to $(\tau, \tau, 0)$.

The third term probes a tensor of rank 2 and is even in time reversal symmetry. This tensor is invarient under the point group symmetry of the unit cell. This term is taken to be proportional to the electric quadrupole moment; i.e. the asphericity of the atomic charge density. The asphericity may be intrinsic to the lattice (as it is in the case of Templeton scattering\cite{templeton:133}) or it can be due to the ordering of electric quadrupole moments.  It is this term which is responsible for the second order satellites observed at $(2\tau, 2\tau, 0)$. These satellites have previously been observed in rare earth systems\cite{gibbs:1241} and due to their appearance at $2\vec{Q}$ positions are normally only observable in non-colinear super-structures. This term may rotate the polarization of the photon. 

\begin{figure}
\includegraphics[width=0.9\columnwidth]{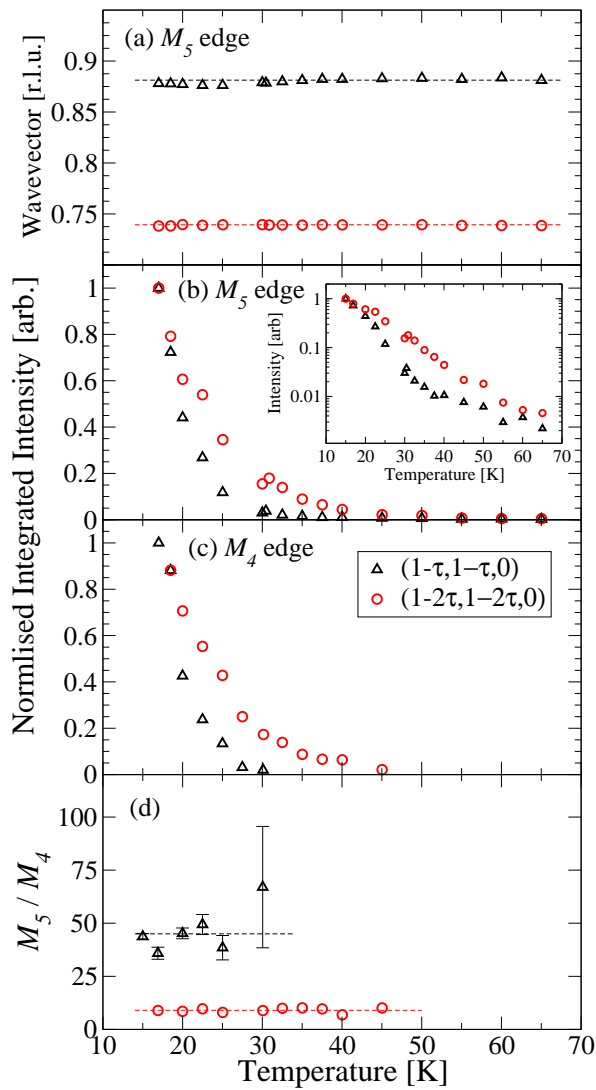}
\caption{\label{dy_edge}(color online) The (a) position and (b) integrated intensity of the \minustau\ and \minustwotau\ superlattice peaks measured at the Dy $M_5$ edge as a function of temperature.  The inset of (b) shows the same plot on a logarithmic scale.  This clarifies the existence of a signal up to 65~K. (c) The temperature dependence of the integrated intensity \minustau\ and \minustwotau\ superlattice peaks as measured at the Dy $M_{4}$ edge.  Panel (d) shows the ratio of the integrated intensity of the superlattice peaks at the $M_5$ and $M_4$ edges.   Note that the integrated intensities in panels (b) and (c) are normalised.  Data measured at X1B.}
\end{figure}

The temperature dependencies of the \minustau\ and \minustwotau\ superlattice peaks at the Dy $M_{4,5}$ edges were measured at the X1B beamline (Fig.~\ref{dy_edge}).    Both superlattice peaks display similar temperature dependences when measured at the $M_5$ edge, and the signal could be detected up to a temperature of 65~K  (Fig.~\ref{dy_edge}(b)). This temperature was slightly greater than that detected through other methods, suggesting soft x-ray diffraction is very sensitive to the dysprosium magnetic moments.   The peaks displayed an exponential decrease in intensity with increasing temperature through the temperature range 17 - 65~K.   Measurements were also taken at the Dy $M_4$ edge, where the signal from the \minustau\ peak was undetectable above 30~K, while the \minustwotau\ peak was observed up to 45~K.

\begin{figure}
\includegraphics[width=0.9\columnwidth]{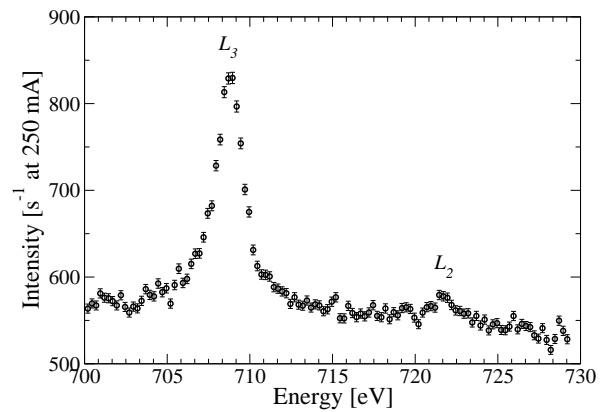}
\caption{\label{fe_res}Scattered x-ray intensity as a function of incident photon energy at the constant wavevector $\vec{Q}=(1-3\tau,1-3\tau,0)$, measured through the Fe $L_{2,3}$ edges. No signal was observed at the Fe $L_1$ edge.   The scan was conducted at 17~K at X1B.}
\end{figure}

\begin{figure}
\includegraphics[width=0.9\columnwidth]{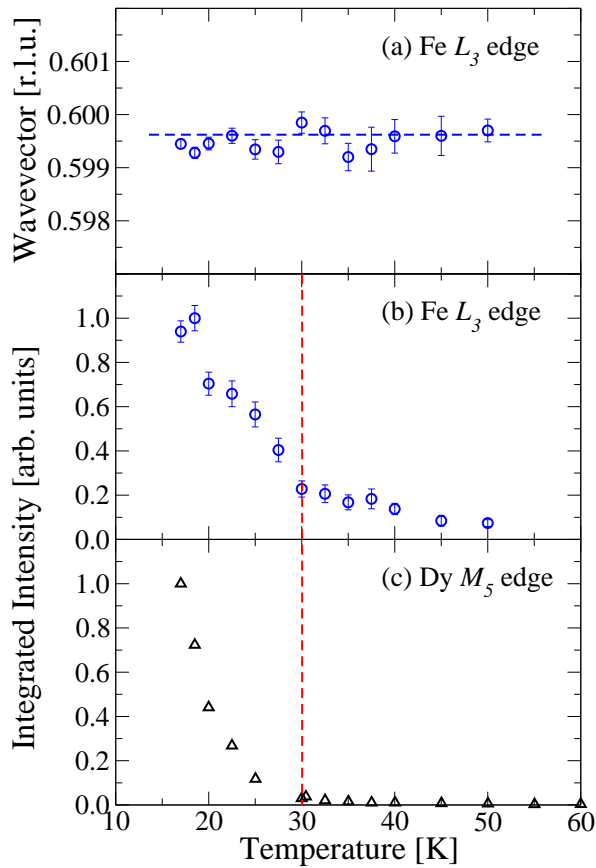}
\caption{\label{fe_edge}(color online) Temperature dependence of the \emph{(a)} wavevector and \emph{(b)} intensity of the \minusthreetau\ peak at the Fe $L_{3}$ edge.  The integrated intensity at the Dy $M_5$ edge as a function of temperature \emph{(c)} is shown for comparison.   Data measured at X1B.  The vertical dashed line is placed at T$_{Dy}$, the horizontal dashed line is a guide to the eye.}
\end{figure}

It has frequently been suggested\cite{talik:361} that the iron moments in these systems orders at $\sim$180~K.   In order to investigate this  we experimented with an incident x-ray energy corresponding to the Fe $L_{2,3}$ edges.  This reduced the size of the Ewald sphere, and as such is was not possible to reach much further out in reciprocal space than the \minusthreetau\ superlattice peak.   A signal from the \minusthreetau\ magnetic peak was seen only at the Fe $L_{2}$ and $L_{3}$ edges (Fig.~\ref{fe_res}), but no signal was observed at the Fe $L_{1}$ edge.  We could not observe a signal from the the \fivetau\ peak.  The resonant spectra of the \minusthreetau\  shows a single peak corresponding to the Fe $L_{3}$ edge, and a very weak peak at the Fe $L_{2}$ edge.  Neither peak displayed any fine structure, when measured with an energy resolution of approximately 220~meV (X1B).  As with the \minustau\ and \minustwotau\ at the Dy $M_5$ edge,  no large changes were observed in the wavevector of the superlattice peak throughout the temperature range (Fig.~\ref{fe_edge}(a)).  The temperature dependence of the \minusthreetau\ superlattice peak is shown in Figure~\ref{fe_edge}(b).     The intensity of the \minusthreetau\ peak was much weaker, even at very low temperatures, than the intensity of the superlattice peaks at the dysprosium absorption edges.   As such, although we were only able to detect the \minusthreetau\ peak as high as 55~K, it does not necessarily follow that the order temperature of the iron is lower than that of the dysprosium.  However, if indeed the ordering of the iron magnetic lattice is as high as 100~K, then these results show  that the ordering of the dysprosium has a large effect on the ordering of the iron.

An alternative explanation for the presence of the \minustwotau\ superlattice peaks, is that they originate from electron density distortions.   Such peaks, also observed by  Langridge \emph{et al.}\cite{langridge:2187}, could originate, as suggested by Langridge \emph{et al} from a magnetoelastic distortion.   If such a crystal lattice distortion existed, this signal would be present through scattering with photons at non-resonant energies.   To test this hypothesis, the sample was mounted on the high energy x-ray diffraction beamline BW5 at HASYLAB, to observe possible weak charge superlattice reflecions.   High energy diffraction is extremely sensitive to small structural distortions, although it is insensitive to magnetic diffraction.  The crystal was aligned and cooled, and despite observing strong Bragg peaks, no superlattice peaks could be seen.  This null result suggests that there is no accompanying structural distortion.   This suggests that the \minustwotau\ peak orginates purely from an aspherical $4f$ electron band, and that the soft x-ray scattering signals are magnetic in origin.

\section{Discussion}

The temperature dependancies of the superlattice peaks at the iron and dysprosium edges show distinctly different behaviour.   Firstly it should be noted that the signal observed at the Fe $L_3$ edge was significantly weaker than that observed at the Dy $M_{5}$ edge.  The signal at the dysprosium edge was first observed at 65~K (Fig. \ref{dy_edge}), and at the iron edge at 55~K (Fig. \ref{fe_edge}).  This initially suggests that the dysprosium moments order before the iron.  However upon cooling, at 30~K the signal at the dysprosium increases substantially, simultaneous to a rather smaller increase in the signal at the iron absorption edge.  Previous studies have consistently detected a relatively high ordering temperature of the iron moment ($\sim100-180$~K), and a much lower ordering temperature of the dysprosium magnetic moment ($\sim30$~K).    Our results therefore suggest that the high temperature ordering of the iron moments induces a small moment on the dysprosium sites.   The relative intensities of the signals observed suggest that the soft energy resonant diffraction is more sensitive to the dysprosium moment, hence the failure to observe the iron moment at higher temperatures.   As the sample is cooled further, the significant increase in the moment on the dysprosium moment corresponds to the `true' ordering of the rare earth.   This is turn effects the size of the magnetic moment on the iron, which is observed to increase simultaneously.   This is in contrast to that suggested by Paix\~{a}o \emph{et al}\cite{paixao:6176}, who suggested from the relative intensities of the (121) (110) peaks, that there is no increase in the iron moment at T$_\mathrm{Dy}$.   Resonant soft x-ray diffraction can detect a relative change in the magnitude of the magnetic moments on the individual atomic species, however it is unable to provide a quantitative measure of the magnetic moment.

The earliest powder neutron measurements\cite{schafer:205} on DyFe$_4$Al$_8$ found only one magnetic transition at 25 K which was attributed to ordering of the iron moments into a conical spiral structure.  It seems likely, however that this transition is in fact due to the ordering of the dysprosium moments.   The suggested ordering temperature for the dysprosium moments of 30~K is supported by temperature dependent measurements of the AC and DC magnetic susceptibilities of DyFe$_4$Al$_8$ by Talik~\emph{et al.}\cite{talik:361}.  They found the magnetic susceptibilities to have a pronounced peak at 30~K.  In addition a thermomagnetic effect was found below 30~K.   Partial ordering of the iron sub-lattice is believed to occur in the temperature region 50-180~K.  Bulk M\"{o}ssbauer measurements also suggest that the iron atoms order at 180~K.  It is difficult from this study to accurately pin down $T_{\mathrm{Fe}}$ as the intensity at the Fe $L_{3}$ edge is very low, and there is no sudden drop in intensity suggesting a sudden transition.   However the behaviour of the signal at the Dy $M_5$ edge is much more conclusive for defining $T_{\mathrm{Dy}}$.

Previous resonant x-ray scattering studies had employed the Dy $L_2$ and $L_3$ edges.   These probe electronic dipole transitions (E1) from the 2$p_{1/2}$ ($L_2$ edge) and the $2p_{3/2}$ ($L_3$ edge) core levels to the empty $5d$ states.   Such techniques are generally only indirectly sensitive to the $4f$ magnetic ordering.  They do however give information on the spin polarisation of the $5d$ band which is coupled to the partially filled $4f$ shell via spin-orbit coupling and band structure effects.  Coupling between the $4f$ moments is important in stabilising many complex magnetic structures via the Ruderman-Kittel-Kasuya-Yosida (RKKY) mechanism.   Systematic studies by Kim \emph{et al.}\cite{kim:064403} of the resonant scattering amplitudes in $R$Ni$_2$Ge$_2$ ($R=$ Gd, Tb, Dy, Ho, Er, Tm) found that the branching ratio (the ratio of the $L_3/L_2$ resonant scatting intensities) varied as a function of the rare earth elements.  In GdNi$_2$Ge$_2$ it the ratio was almost unity, then increased to 3.5 in DyNi$_2$Ge$_2$ and then to 8.0 in ErNi$_2$Ge$_2$.   Such effects are due to increasing spin-orbit coupling and $4f$ orbital polarisation.   In nearly all resonant x-ray scattering studies of particular materials this branching ratio is constant at all temperatures.  This is because the electronic band polarisation effects are generally insensitive to thermal effects.

Langridge \emph{et al.}\cite{langridge:2187} found that in DyFe$_4$Al$_8$ this branching ratio ($L_3/L_2$) was temperature dependent.  Below 20~K the Dy $L_3$ resonance dramatically increased in intensity whereas the Dy $L_2$ resonance decreased.   In all these measurements the resonances were very simple, comprised of a single E1 dipole transition.   There was no sign of an higher multipole orders such as E2 (electric quadrupole).  Langridge \emph{et al.} speculated that the temperature dependence of the branching ratio was caused by changes in the $5d$ band polarisation caused by the increasing ordered component of the Dy $4f$ moment and simultaneous magnetoelastic distortions below $T_{\mathrm{Dy}}$.  Our measurements (Fig.~\ref{dy_edge}(d)) show no change in the branching ratio of the integrated intensity a the $M_5$ and $M_4$ edges).   It is clear that even if there are changes in DyFe$_4$Al$_8$ below 30~K in the spin polarisation of the $5d$ band, these are not replicated in the $4f$ band.

It was noted by Paix\~{a}o~\emph{et al.}\cite{paixao:6176} that the modulation vector of the magnetic superlattice changes slightly below T$_N$.    Although no significant change is obvious in our results (Fig~\ref{dy_edge}(a)), there is a slight variation from a constant value which is compatible with the previously published results.

\section{Conclusion}

The magnetic structure of \dyfeal\ has been studied using resonant soft x-ray diffraction.  Superlattice peaks have been observed at the Dy $M_{4,5}$ edges and Fe $L_{2,3}$ edges.   These have shown different temperature dependencies suggesting the iron initially orders at $T_{\mathrm{Fe}}\geq65$~K, and the dysprosium ordering at $T_{\mathrm{Dy}}\approx30$~K.  The ordering of the dysprosium appears to be significantly stronger, although quantitative comparisons are difficult to make.   The superlattice peaks at the Dy $M_5$ edge appear at $(1\pm n\tau,1\pm n\tau,0)$, where $n$ is both odd and even.   It is unclear what the origin of the peaks at even $n$ are due to.   There was no observation of a change in wavevector of any of the superlattice peaks throughout the temperature range that they could be observed.

\section{Acknowledgements}
TAWB wishes to thank EPSRC for support. PDH thanks the University of Durham Research Foundation for support.  PDH and TAWB acknowledge travel assistance from EPSRC.  The authors would like to acknowledge the European Synchrotron Radiation Facility for provision of synchrotron radiation facilities.  X-ray measurements at X1B, NSLS were supported by the Office of Basic Energy Sciences, U.S. Department of Energy under Grant No. DE-FG02-06ER46285, with use of the NSLS supported under Contract No. DE-AC02-98CH10886.    \bibliography{dyfeal}

\end{document}